\documentclass[twocolumn,showpacs,preprintnumbers,amsmath,amssymb,prl]{revtex4}

\usepackage{graphicx}
\usepackage{dcolumn}
\usepackage{bm}

\begin{document}

\input epsf.sty

\widetext

\title
{
Gigantic Impact of Magnetic Impurity on Stripe Order in La$_{2-x}$Sr$_{x}$CuO$_4$ ($x$$\sim$1/8)
}

\author{M. Fujita$^{1}$}
\email{fujita@imr.tohoku.ac.jp}
\author{M. Enoki$^{2}$}
\author{S. Iikubo$^{3}$}
\author{K. Kudo$^{1}$}
\author{N. Kobayashi$^{1}$}
\author{K. Yamada$^{3}$}

\affiliation{%
$^{1}$ Institute for Materials Research, Tohoku University, Sendai, Miyagi 980-8577, Japan\\
$^{2}$ Department of Physics, Tohoku Univsersity, Sendai, Miyagi 980-8578, Japan\\
$^{3}$ World-Premier-International Research Center Initiative, Tohoku University, Sendai, Miyagi 980-8577, Japan
}%

\date{\today}




\begin{abstract}

Precise neutron-scattering experiments on La$_{1.88-y}$Sr$_{0.12+y}$Cu$_{0.99}$R$_{0.01}$O$_4$ ($y$=0 for R=Cu, Zn and $y$=0.01 for R=Ga, Fe) were performed to examine the stability of stripe order in cuprate oxides through impurity substitution effects. For all the impurity-substituted samples a static charge-density-wave order similar to that reported in the low-temperature-tetragonal phase (LTT) was discovered in the low-temperature-orthorhombic (LTO) phase for the first time. Fe substitution significantly enhanced both the charge- and spin-density-wave orders. These results indicate that dynamical charge stripes can be stabilized into static ones around the Cu-site defects regardless of the type of crystal structure. Additionally, large magnetic moments at the Cu-site may induce a decrease in the Cu spin fluctuations around the defects, which in turn, enhances the static charge-density-wave order. 
\end{abstract}



\maketitle

In lamellar copper oxides, the evidence of an interplay between spin and charge degree of freedom is increasing. Hole doping into an antiferromagnetic (AF) insulator of La$_2$CuO$_4$ rapidly suppresses the N\'{e}el temperature ($T_{\rm N}$) while simultaneously altering the electrical transport properties~\cite{Kastner_98}. On the other hand, the $T_{\rm N}$ for the doped antiferromagnet can be recovered by impurity doping onto the CuO$_2$ planes as the electrical resistivity increases~\cite{Hucker_99}. Therefore, the stability of the AF order and localization of holes is strongly correlated. Furthermore, a similar enhancement of incommensurate (IC) AF order has been observed in La$_{2-x}$Sr$_x$CuO$_4$ (LSCO) at $x$$\sim$1/8 by either impurity doping~\cite{Kimura_03} or applying magnetic fields~\cite{Katano_00,Lake_02}. The origin of such enhanced IC AF order, i.e. the spin-density-wave (SDW) order, has been intensively studied due to the close connection with the mechanism of high-$T_{\rm c}$ superconductivity~\cite{Sachdev_03}. 

Static stripe formation of doped holes with segregating one-dimensional spin domains has been observed in La$_{1.6-x}$Nd$_{0.4}$Sr$_x$CuO$_4$ (LNSCO)~\cite{Tranquada_95} and La$_{2-x}$Ba$_{x}$CuO$_4$ (LBCO)~\cite{Fujita_04} with $x$$\sim$1/8. The stripe order, which is the prominent phenomenon caused by competition between spin and charge degree of freedom, is statically stabilized by pinning or decreased dynamical fluctuation, which leads to the suppression of the competing superconductivity. To date, superlattice peaks corresponding to the charge-density-wave (CDW) order have been observed only in the low-temperature tetragonal phase (LTT, \textit{P}4$_2$/\textit{ncm} symmetry)~\cite{Fujita_02,Kimura_04}. Experimental evidence for the static CDW order in the low-temperature orthorhombic phase (LTO, \textit{Bmab} symmetry) has yet to be reported, but the SDW order in LSCO can be strengthened by impurity substitution or under an external magnetic field. These experimental facts suggest that SDW and CDW have contrasting responses to defects and magnetic fields. Therefore, the universality of stripe correlations and the intrinsic charge distribution on CuO$_2$ planes are debatable.  

To gain further insight into above problem, we performed precise and comprehensive neutron-scattering measurements on the magnetic (Fe$^{+3}$) and non-magnetic (Zn$^{+2}$ and Ga$^{+3}$) impurity-doped as well as pristine LSCO systems with effective hole concentrations of 0.12. The static CDW order is induced by a small amount of impurity even in the LTO phase, implying that regardless of the crystal structure and doped impurity a static stripe order can be realized. Additionally, the CDW order is remarkably pronounced in the Fe-substituted sample (Fe-LSCO). Similarly, Fe doping strongly increases the intensity of the SDW peak, although the effects in Zn-doped (Zn-LSCO) and Ga-doped (Ga-LSCO) systems are negligible. Therefore, the stability of static stripe order is effectively enhanced by magnetic impurities rather than non-magnetic ones or charge impurities with different Cu valences. 

For the experiments, we grew single crystals of a pristine LSCO system and those doped with either Fe, Zn, or Ga by the traveling-solvent floating-zone method. In contrast to Zn$^{2+}$ (3d$^{10}$, S=0), Fe and Ga ions have been reported to be doped as higher valences of Fe$^{3+}$ (3d$^{5}$, S=5/2) and Ga$^{3+}$ (3d$^{10}$, S=0), which result in a reduction of one hole by one doped ion. Hence, to study the intrinsic impurity effect on the stripe order, we used the La$_{1.88-y}$Sr$_{0.12+y}$Cu$_{0.99}$R$_{0.01}$O$_4$ ($y$=0 for R=Cu, Zn and $y$=0.01 for R=Ga, Fe) system where the amount of impurity and hole concentration are fixed at 0.01 and 0.12, respectively. Neutron-scattering experiments were performed on the Tohoku University triple-axis spectrometer, TOPAN at the JRR-3 reactor of the Japan Atomic Energy Agency (JAEA). We set the incident neutron energy $E_{\rm i}$ to 14.7 meV using the (002) pyrolytic graphite analyzer reflection. In a typical measurement of CDW(SDW) order, the collimator sequences of 30$^{\prime}$(15$^{\prime}$)-30$^{\prime}$-30$^{\prime}$-180$^{\prime}$ with momentum and energy resolutions of $\sim$0.05$\AA^{-1}$ and $\sim$2.4 meV (full-width-at-half-maximum) respectively was selected. For each systems, a couple of cylindrical crystals with diameters of 8 mm and lengths 35 mm were assembled with the ({\it h} {\it k} 0) plane parallel to the scattering plane. Then the samples were mounted in a closed-cycle refrigerator to control the temperature. With this sufficient sample volume, we successfully observed the impurity-induced CDW order by a neutron-scattering measurement. For convenience, herein we use a tetragonal notation to describe the crystallographic indices (1 r.l.u.=1.67 $\AA^{-1}$)~\cite{lattice}.

\begin{figure}[t]
\begin{center}
\epsfxsize=2.75in\epsfbox{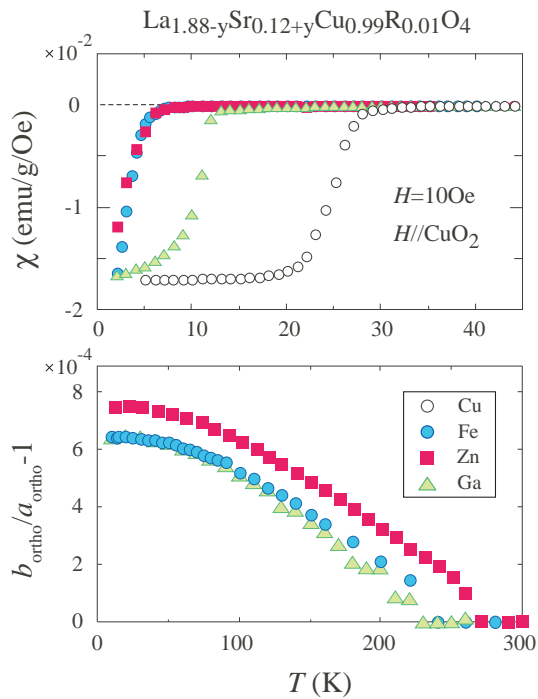}
\caption
{(Color online): 
Temperature-dependence of (a) zero-field-cooled magnetic susceptibility $\chi$ measured at 10 Oe and (b) orthorhombicity defined as $b_{\rm ortho}$/$a_{\rm ortho}$-1 for La$_{1.88-y}$Sr$_{0.12+y}$Cu$_{0.99}$R$_{0.01}$O$_4$ with $y$=0 for Cu and Zn, and $y$=0.01 for Fe and Ga.} 
\end{center}
\end{figure}

Figure 1(a) shows the temperature dependence of the magnetic susceptibility measured for the all samples. Impurity substitution reduced the superconducting transition temperature ($T_{\rm c}$). The $T_{\rm c}$$^{\prime}$s values, which are defined as the temperature at where the susceptibility is 5$\%$ of that at lowest temperature measured, are 5.7 K, 6.1K, 11.6 K, and 27.9 K for Fe-, Zn- and Ga-doped, and pristine samples, respectively. These values are almost consistent with the previously reported results for powder samples~\cite{Xiao_90, Gaojie_00}.

\begin{figure}[t]
\begin{center}
\epsfxsize=2.7in\epsfbox{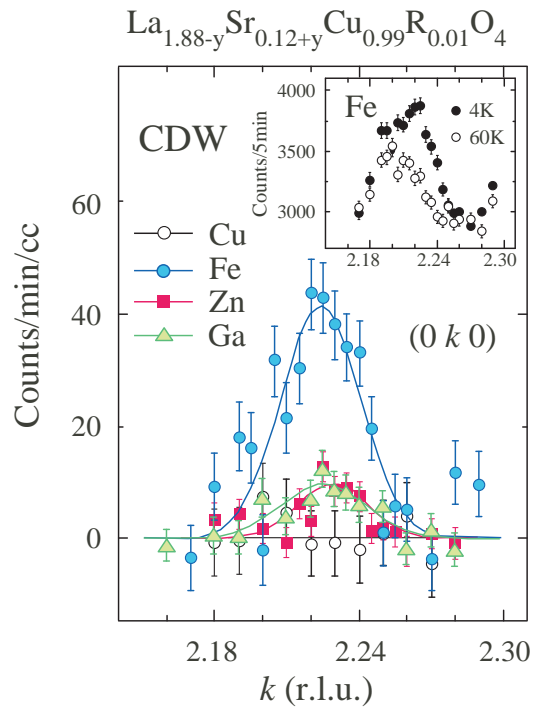}
\caption
{(Color online): 
Remnant intensity from CDW order obtained after subtracting the high-temperature ($T$$\geqslant$60 K) data from the low-temperature ($T$$\leqslant$10 K) data for La$_{1.88-y}$Sr$_{0.12+y}$Cu$_{0.99}$R$_{0.01}$O$_4$. Inset shows the observed incommensurate peak in La$_{1.88}$Sr$_{0.12}$Cu$_{0.99}$Fe$_{0.01}$O$_4$.} 
\end{center}
\end{figure}

To confirm the absence of the LTT phase at low temperature, we measured the temperature dependence of the intensity at the allowed superlattice position of (100) in the LTT phase and the orthorhombic $a$- and $b$-lattice constants with high-\mbox{\boldmath $Q$} resolution at the KINKEN triple-axis spectrometer, AKANE. In all samples, as expected for the LTO structure, the intensity at the (100) position did not show a temperature-dependence within the experimental accuracy. Furthermore, as seen in Fig. 2(b), the orthorhombicity (=$b_{\rm ortho}/$a$_{\rm ortho}$-1) monotonously decreased as the temperature increased, suggesting the absence of a structural transition between LTO and LTT. From these results, we concluded that the small amount of impurity substitution does not fundamentally alter the bulk crystal structure and the LTO structure remains in the La$_{1.88-y}$Sr$_{0.12+y}$Cu$_{0.99}$R$_{0.01}$O$_4$ system. Both the onset temperature for the appearance of LTO phase and the orthorhombicity at low temperature in impurity-doped LSCO were consistent with those for pristine LSCO with an equivalent Sr concentration, suggesting a negligible effect of impurity doping on the bulk crystal structure.

\begin{figure}[t]
\begin{center}
\epsfxsize=2.7in\epsfbox{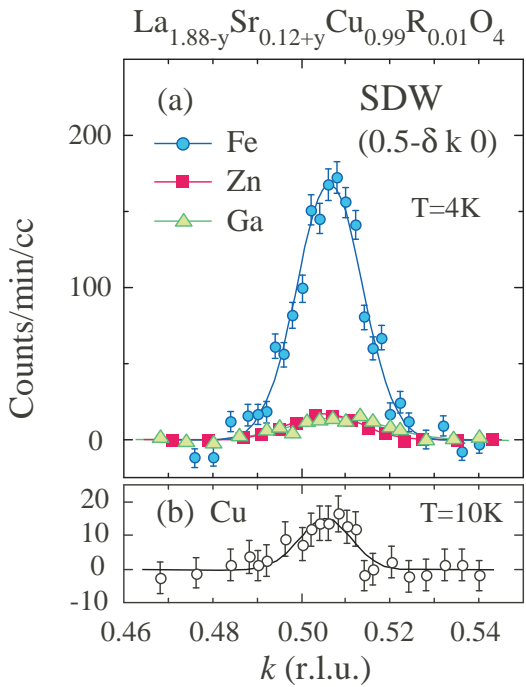}
\caption
{(Color online): 
Incommensurate peak from the SDW order in (a) impurity-doped La$_{1.88-y}$Sr$_{0.12+y}$Cu$_{0.99}$R$_{0.01}$O$_4$ with R=Fe, Zn and Ga, and (b) pristine La$_{1.88}$Sr$_{0.12}$CuO$_4$. Profiles are scanned along [0.5-$\delta$ $k$ 0] in the reciprocal space where $\delta$$\sim$0.115. Intensities are normalized by the sample volume and the counting time after subtracting the background slope. 
}
\end{center}
\end{figure}

Figure 2 depicts the remnant intensity of the CDW peak, which is normalized by the sample volume and counting time after subtracting the high-temperature ($T$$\geqslant$60 K) data from the low-temperature ($T$$\leqslant$10 K) data. The inset shows a representative spectrum measured for Fe-LSCO. Because all the measurements were performed using the same spectrometer setup, the normalized intensity directly corresponds to the strength of the CDW order. In the pristine sample, clear evidence of CDW order was not observed, which is consistent with previous reports. However, the IC superlattice peaks induced by the CDW order were observed in all the impurity-doped samples. Thus, regardless of the type of crystal structure, stripe correlations potentially exist in the La-214 system, and the CDW order can be stabilized by doping with a small amount of impurity on the CuO$_2$ planes. 
However, the intensity of the CDW peak was much stronger in Fe-LSCO than those the Zn and Ga-LSCO. To quantitatively analyze the result for Fe-LSCO, a two-dimensional Gaussian function was fitted to the resultant intensity after considering the experimental resolution. The evaluated peak center and the resolution-corrected coherence length for the lattice distortion $\xi$$_{\rm CDW}$, which is the inverse of the intrinsic peak-width, were (0 2+$\epsilon$ 0) with $\epsilon$=0.224(2) and 60(5) $\AA$, respectively. $\xi$$_{\rm CDW}$($\sim$60 $\AA$) exceeded the mean distance between the nearest neighbor Fe ions ($\sim$38 $\AA$), suggesting that a bulk CDW order is realized and coexists with the LTO phase. Although $\xi$$_{\rm CDW}$ in Fe-LSCO is somewhat shorter than that in La$_{1.875}$Ba$_{0.125}$CuO$_4$ ($\xi$$_{\rm CDW}$$\sim$100$\AA$), the integrated peak intensity and the peak position in the two systems were comparable. 

\begin{figure}[t]
\begin{center}
\epsfxsize=2.8in\epsfbox{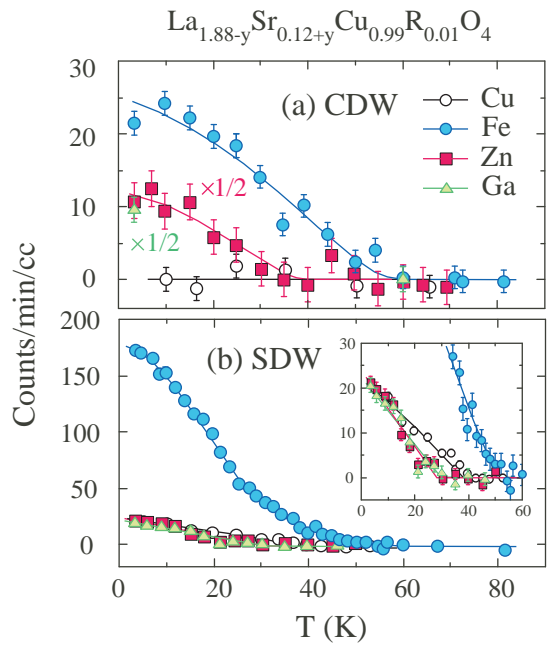}
\caption
{(Color online): 
Temperature dependences of volume-corrected superlattice peak intensities for (a) CDW and (b) SDW orders in La$_{1.88-y}$Sr$_{0.12+y}$Cu$_{0.99}$R$_{0.01}$O$_4$ with R=Cu, Fe, Zn and Ga. The CDW intensity for Zn(Ga)-LSCO is scaled by a factor of two. 
}
\end{center}
\end{figure}

Figure 3 shows similar profiles for the SDW order peaks scanned around ($\pi$, $\pi$) reciprocal position for (a) impurity-doped and (b) pristine systems. Well-defined IC peaks from SDW order were observed in all samples with an anisotropic peak-shift from the high symmetric position ($k$=0.5), which corresponds to a deviation of the SDW wave vector from the Cu-O bond direction~\cite{YoungLee_99, Kimura_00, Fujita_02_2}. The normalized intensity of the SDW peak was much stronger in Fe-LSCO, analogous to the case for CDW order. Compared to the pristine LSCO, Fe doping increased the integrated intensity in the ($HK$0) momentum plane by a factor of $\sim$8, but clear enhancements were not detected in the cases of Zn and Ga dopings. On the other hand, the incommensurability $\delta$ defined as half the distance between a pair of IC peaks to ($\pi$, $\pi$) had a similar value of 0.115(3) r.l.u. for all samples, and the magnitude of the peak shift was not influenced by the impurity substitution. Thus, impurity doping does not introduce a significant effect on both the modulated direction and the periodicity of SDW order.  However, Fe doping effectively stabilizes the existing SDW order. The correlation length ($\xi$$_{\rm SDW}$) evaluated after the resolution collection procedure was $\sim$130(20) $\AA$ for the impurity-doped LSCO. Because the $\xi$$_{\rm SDW}$ for the pristine sample has been reported to exceed $\sim$200 $\AA$ from a high-\mbox{\boldmath $Q$} resolution measurement~\cite{Kimura_99}, the $\xi$$_{\rm SDW}$ was reduced by a similar extent for doping all impurities. 

Figure 4 represents the temperature dependence of the volume-corrected intensity for the IC peaks from (a) CDW and (b) SDW orders. For each samples, the peak intensity was monitored by the temperature change. Even in the LTO phase, the CDW order appeared at $T_{\rm CDW}$ of $\sim$60 K and $\sim$40 K in Fe-LSCO and Zn-LSCO, respectively, which are comparable to $T_{\rm CDW}$ of 50K in the LTT phase of LBCO. In contrast, the magnetic intensity showed a different impurity effect between magnetic and non-mangetic doped ions. That is, the intensity was strongly enhanced by Fe doping as the magnetic ordering temperature $T_{\rm SDW}$ increased, but both the intensity and $T_{\rm SDW}$ exhibited a negligible change or slight reduction by Zn and Ga doping within the experimental accuracy. (See inset) It should be noted that the effects of Zn and Ga doping on the CDW and SDW orders were similar, regardless of their different valence states. 

What is the microscopic origin for the impurity-induced static CDW, particularly the gigantic effect of Fe impurity? Generally, the origin is strongly related to carrier localization due to impurity-induced randomness on CuO$_2$ planes either by charge valence, lattice distortion, or spin defects. For example, the substantial difference in the ionic radius between Cu$^{2+}$ (0.73 $\AA$) and Ga$^{3+}$ (0.62 $\AA$) or Fe$^{3+}$ (0.645 $\AA$) might induce a local LTT-type distortion and the static CDW order as seen in the LTT structure in Ba-doped La$_2$CuO$_4$. However, it is difficult for Zn to induce such a distortion because both ionic radius and valence state (Zn$^{2+}$ (0.74 $\AA$)) are nearly the same as Cu. Rather than a structural effect, the similar effects on the CDW and SDW orders between Zn and Ga strongly suggest spin defects play the dominant role.

We emphasize that it is difficult to understand the gigantic effect of Fe impurity only from the randomness effect discussed above. Clearly a giant spin-effect exists for the Fe impurity.  We hypothesize that the large spin of the Fe impurity suppresses the dynamical fluctuation of the surrounding Cu spins. In fact, the volume averaged magnetic moment of 0.3 $\mu$$_{\rm B}$ for the present Fe-LSCO is much larger than the values from the doped Fe spins (1$\%$ of 5 $\mu$$_{\rm B}$ = 0.05 $\mu$$_{\rm B}$) and the ordered moment of Cu spins in the pristine LSCO ($\sim$0.1 $\mu$$_{\rm B}$)~\cite{Kimura_99}. Therefore, the Fe spins induce static or slowly fluctuated Cu spins around Fe. Because spin fluctuations strongly couple with the hole mobility in this system, suppressing spin fluctuations reduces hole mobility, which results in the development of static CDW orders. The ordered moment in the CDW ordered LTT phase of LBCO is larger than that in the LTO phase of LSCO without the CDW order~\cite{Fujita_02}. These facts clearly demonstrate a strong coupling between carrier localization and suppression of spin fluctuations in this system.  

The present results on the impurity effect support a Swiss cheese model proposed from muon spin rotation measurement on Zn-LSCO ~\cite{Nachumi_98, Adachi_08} where the non-SC region is formed around impurities and local moments appear inside the region. We are the first to find that the induced local moment is much larger in the Fe-doped system. Although Fe greatly impacts the CDW and SDW orders, the reduction rate of $T_{\rm c}$ is comparable regardless of the impurity. We speculate that this fact correlates with the comparable size of the static SDW order around impurities. (See the SDW peak width in Fig. 3.) 

In summary, the impurity effect on the static stripe order in the LSCO system near the 1/8 doping was comprehensively studied by neutron-scattering measurements. We found direct evidence of an impurity-induced static CDW order in the LTO phase.  Moreover, a magnetic Fe impurity remarkably stabilized the spin and charge stripe orders. The intensity of the incommensurate peak from the SDW order was significantly enhanced by Fe$^{3+}$ doping, whereas Zn$^{2+}$ and Ga$^{3+}$ doping shows negligible effect on the intensity of this peak. Therefore, although any defect could induce CDW, the stripe order was more stabilized by a magnetic impurity. These results suggest that an effective approach for exploring the stripe order in other cuprare oxides such as YBa$_2$Cu$_3$O$_{6+\delta}$ and Bi$_2$Sr$_2$Ca$_{n-1}$Cu$_n$O$_{4+2n+\delta}$ is to investigate the impurity effect, especially a magnetic one with large moment.  

We greatly thank to K. Iwasa for sharing the machinetime of neutron scattering measurement at TOPAN, and to Adachi, Y.J. Kim, S.H. Lee, B. Fine, O. Sushkov, J.M. Tranquada, and H. Yamase for stimulating discussions. We also thank to M. Sakurai for supporting the crystal growth and K. Nemoto for the assistance of neutron-scattering experiment at JAEA. This work was supported in part by Scientific Research on Priority Area, (S), (B) (19340090) and (C) (20540342) from the Japanese Ministry of Education, Culture, Sports, Science and Technology.


\begin{references}
\bibitem{lattice} The orthorhombic in-plane lattice constants of $y$=0.01 samples, for instance, are a=5.328$\AA$ and b=5.362$\AA$ at 4 K. The tetragonal lattice constant is approximately 1/$\sqrt{2}$ times the tetragonal one.
\end{references}
\end{document}